\documentclass[runningheads]{svmult}

\usepackage{makeidx}   % allows index generation
\usepackage{graphicx}  % standard LaTeX graphics tool
\usepackage{subeqnar}  % subnumbers individual equations
\usepackage{multicol}  % used for the two-column index
\usepackage{physprbb}  % modified textarea for proceedings,
\makeindex             % used for the subject index
\begin{document}
\title*{Relative Ages of Globular Clusters}
\toctitle{Relative Ages of Globular Clusters}
%\protect\newline in the Particle Deflection Plane}
% allows explicit linebreak for the table of content
%
%
\titlerunning{Relative Ages of Globular Clusters}
% allows abbreviation of title, if the full title is too long
% to fit in the running head
%
\author{Thomas H. Puzia\inst{1}}
%\and Elsa Bertino\inst{1}}
%
\authorrunning{Thomas H. Puzia}
% if there are more than two authors,
% please abbreviate author list for running head
%
%
\institute{Sternwarte der Universit\"at M\"unchen,
	Scheinerstrasse 1, D-81679 M\"unchen, Germany}

\maketitle              % typesets the title of the contribution

\begin{abstract}
Ages of extragalactic globular clusters can provide valuable insights
into the formation and evolution of galaxies. In this contribution the
photometric methods of age dating old globular cluster systems are
summarised. The spectroscopic approach is reviewed with an emphasis of
the right choice of age diagnostics. We present a new method of
quantifying the relatively best age-sensitive spectroscopic index
given the quality of a data set and a certain theoretical stellar
synthesis model. The relatively best diagnostic plot is constructed
from the set of Lick indices and used to age date globular clusters in
several early-type galaxies which are part of a large spectroscopic
survey of extragalactic globular cluster systems. We find that,
independently of host galaxy, metal-poor ([Fe/H]$<-0.8$) globular
clusters appear to be old ($t> 8$ Gyr) and coeval. Metal-rich clusters
show a wide range of ages from $\sim 15$ down to a few Gyr.
\end{abstract}

%%%%%%%%%%%%%%%%%%%%%%%%%%%%%%%%%%%%%%%%%%%%%%%%%%%%%%%%%%%%%%%%%%%%%%%%%%%
\section{Introduction}

The foundation of understanding galaxy formation and evolution is the
access to accurate ages of stellar populations which build up
galaxies. All galaxies are expected to host multiple stellar
populations with different ages, but it is uncertain when and how they
formed. In general, the ages of galaxies are determined from
photometric and/or spectroscopic parameters which are more or less
correlated with the mean lifetime of the observed stellar
system. Looking at the diffuse light of galaxies one observes the
luminosity-weighted sum of all contributions emitted by a mix of
stellar populations with various ages. Hence, it is meaningless to
assign a single age to galaxies and impossible to resolve the age
spread of the underlying stellar populations by looking at the diffuse
light alone.

In the recent past, globular clusters increasingly played the role of
accurate age tracers. Globular clusters are simple stellar populations
(SSPs, containing stars of same age and same metallicity), are formed
during major star formation events, and are easy to observe out to
large distances. The easy age dating of these witnesses of galaxy
evolution makes it possible to assign single starburst events to
globular cluster sub-populations and thus reconstruct the star
formation histories of galaxies. However, there are still fundamental
limits of accurate age dating: (1) photometric and spectroscopic age
diagnostics suffer from the well-known age-metallicity degeneracy and
(2) stochastic fluctuations in bright stellar evolutionary phases in
less-massive clusters ($\leq 10^5 M_\odot$) limit the accuracy of age
determinations. The remedy for the former point is a smart choice of
photometric colours and/or spectroscopic indices to reduce the
age-metallicity degeneracy. The latter problem can be by-passed by
averaging over a large sample of globular clusters.

In the following, the two fundamental techniques of age dating of old
globular clusters are discussed. We will focus on globular cluster
systems which are older than 1 Gyr and hosted in early-type
galaxies. Finally, first results of our spectroscopic survey of
extragalactic globular cluster systems are presented.

%%%%%%%%%%%%%%%%%%%%%%%%%%%%%%%%%%%%%%%%%%%%%%%%%%%%%%%%%%%%%%%%%%%%%%%%%%%
\section{Photometric Age Dating}

\begin{figure}[!t]
\begin{center}
\includegraphics[bb=18 430 592 718,clip,width=12cm]{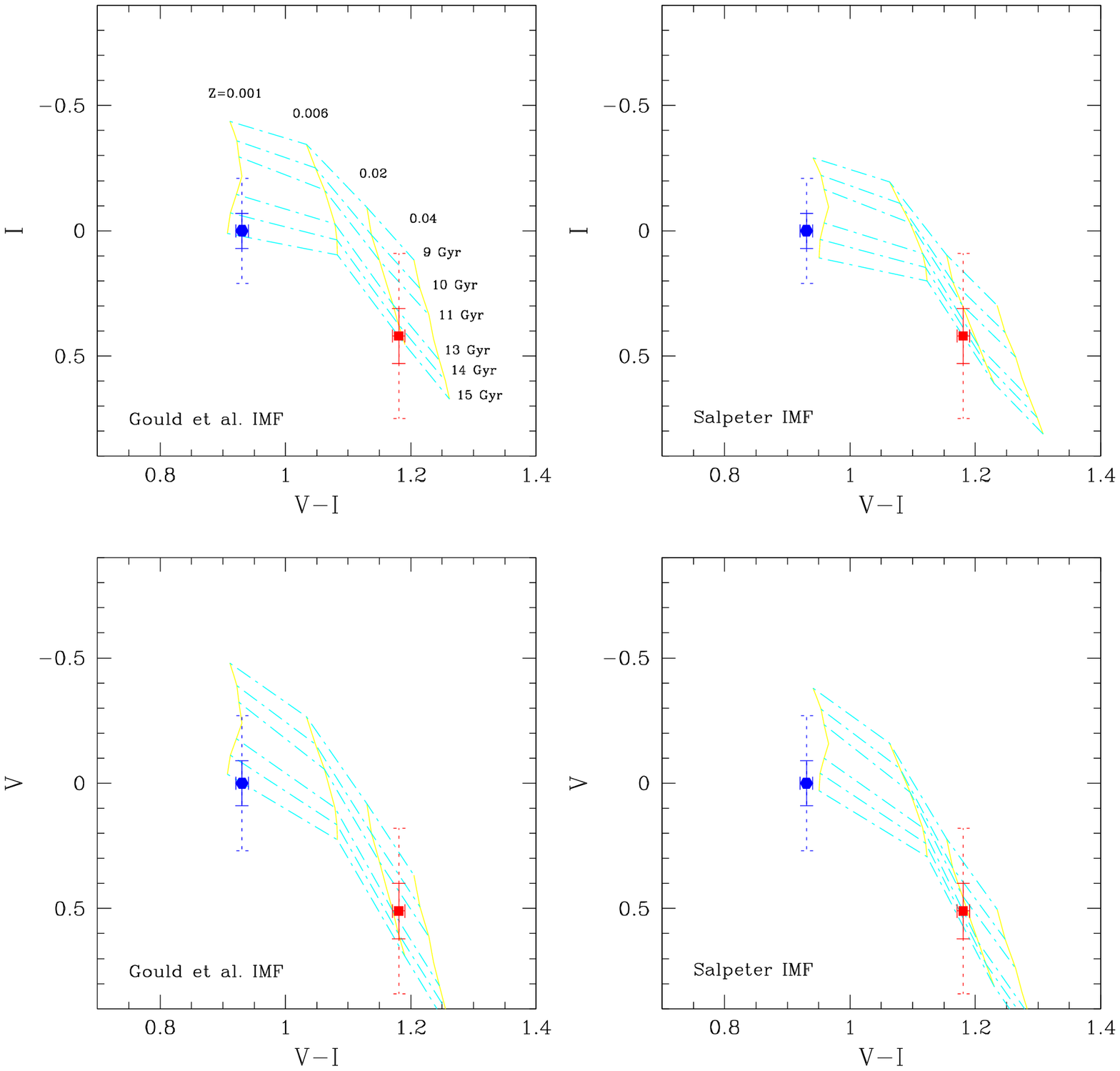}
\end{center}
\caption[]{Age dating of globular clusters in NGC~4472 taken from
\cite{puzia99}. The dots indicate the colours and turn-over magnitudes
of the two major globular cluster sub-populations which are compared
to the SSP model predictions. Different CMD and SSP models variations
were analysed and gave very similar results. The two globular cluster
sub-populations were found to differ mainly in metallicity. For
details see \cite{puzia99}}
\label{ps1}
\end{figure}

Photometric ages of extragalactic globular clusters are obtained from
the comparison of observed parameters with theoretical model
predictions. Such models are calibrated on Local Group globular
clusters for which the ages are known from other techniques. However,
the globular clusters in the Local Group are not always representative
for the variety of clusters found outside the Local Group, e.g. in
early-type galaxies. To obtain meaningful results for clusters with
high metallicities which cannot be observed locally one uses
extrapolations to derive their ages. Absolute age predictions vary
among SSP models as they depend on the ingredients of the model and
the choice of local calibrators. It is essential to use more than one
model when accurate absolute ages are required to secure oneself from
model-to-model variations. However, relative ages can be reliably
obtained using any SSP model as the differential predictions are less
affected by the choice of a specific model. Photometric age dating
methods use colour-magnitude diagrams (CMDs) and colour-colour
diagrams as their tools.

\subsection{Colour-Magnitude Method}
The first method is based on mean colours and turn-over magnitudes of
major globular cluster sub-populations. The requirement is that the
sub-populations can be separated by their colours and their luminosity
function turn-over magnitudes can be observed. In other words, this
method is only applicable to old globular cluster systems with
``evolved'' globular cluster luminosity functions. An application is
shown in Figure \ref{ps1} where relative ages of the two major
globular cluster sub-populations in NGC~4472 can be read off directly
from the model grid. The two sub-populations were found to be coeval
within the errors and to differ mostly in metallicity. The use of
other photometric filters can enhance the predictive power of this
method (see contribution of Andr\'{e}s Jord\'{a}n in this volume). The
maximum age resolution of this method at old ages ($\geq 10$ Gyr) can
reach $2-3$ Gyr and is strongly dependent on the photometric
quality. This technique, however, is premised on the assumption that
all sub-populations feature the same globular cluster mass
function. On the other hand, if extended to a wider wavelength range
this technique can be used to study variations in the globular cluster
mass function. For example, accurate $V$ and $I$ photometry in
NGC~4472 already allowed to constrain the variations in the mass
characterizing the turn-over to be less than 20\%.

\begin{figure}[!t]
\begin{center}
\includegraphics[width=6cm]{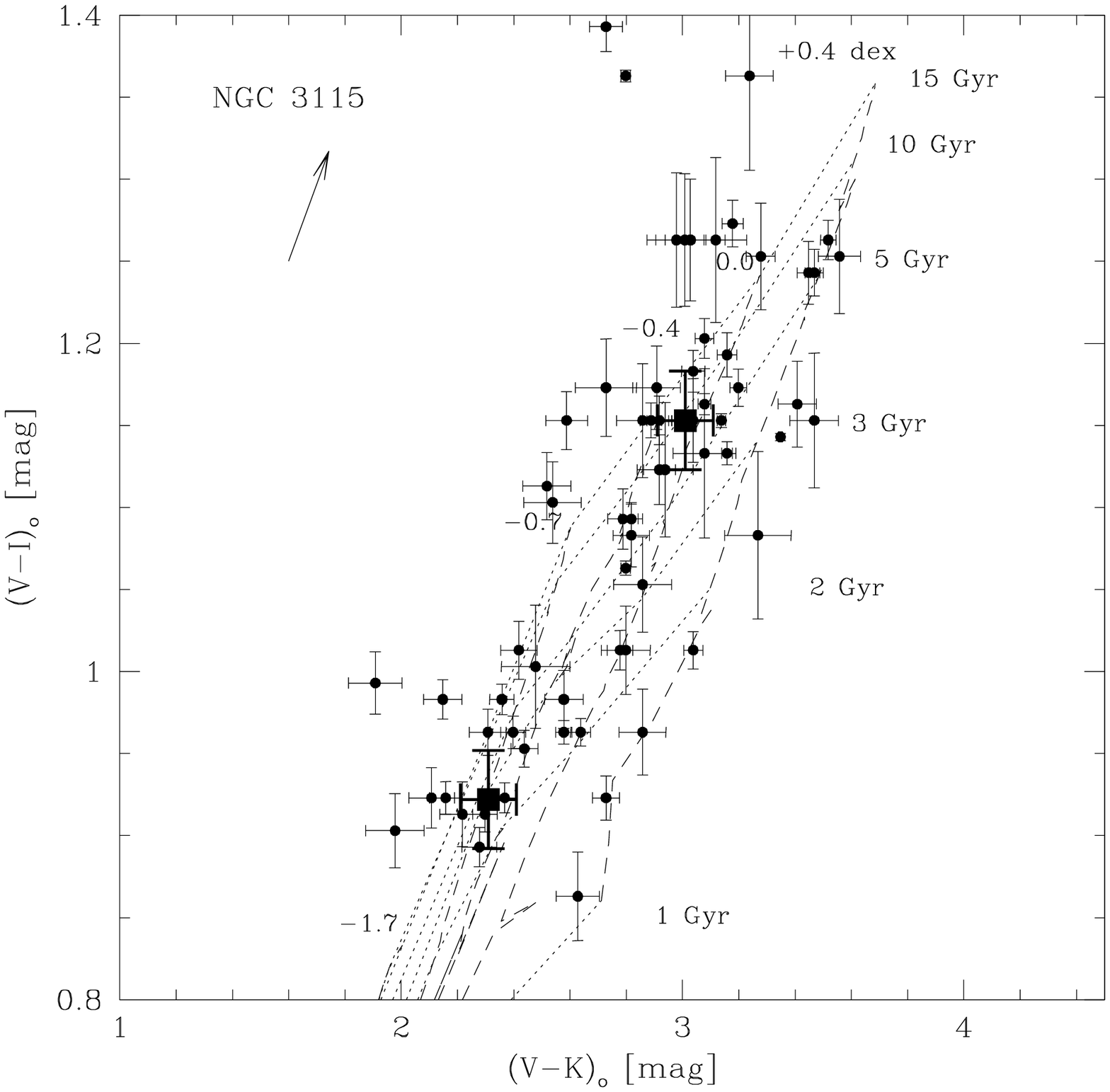}
\includegraphics[width=6cm]{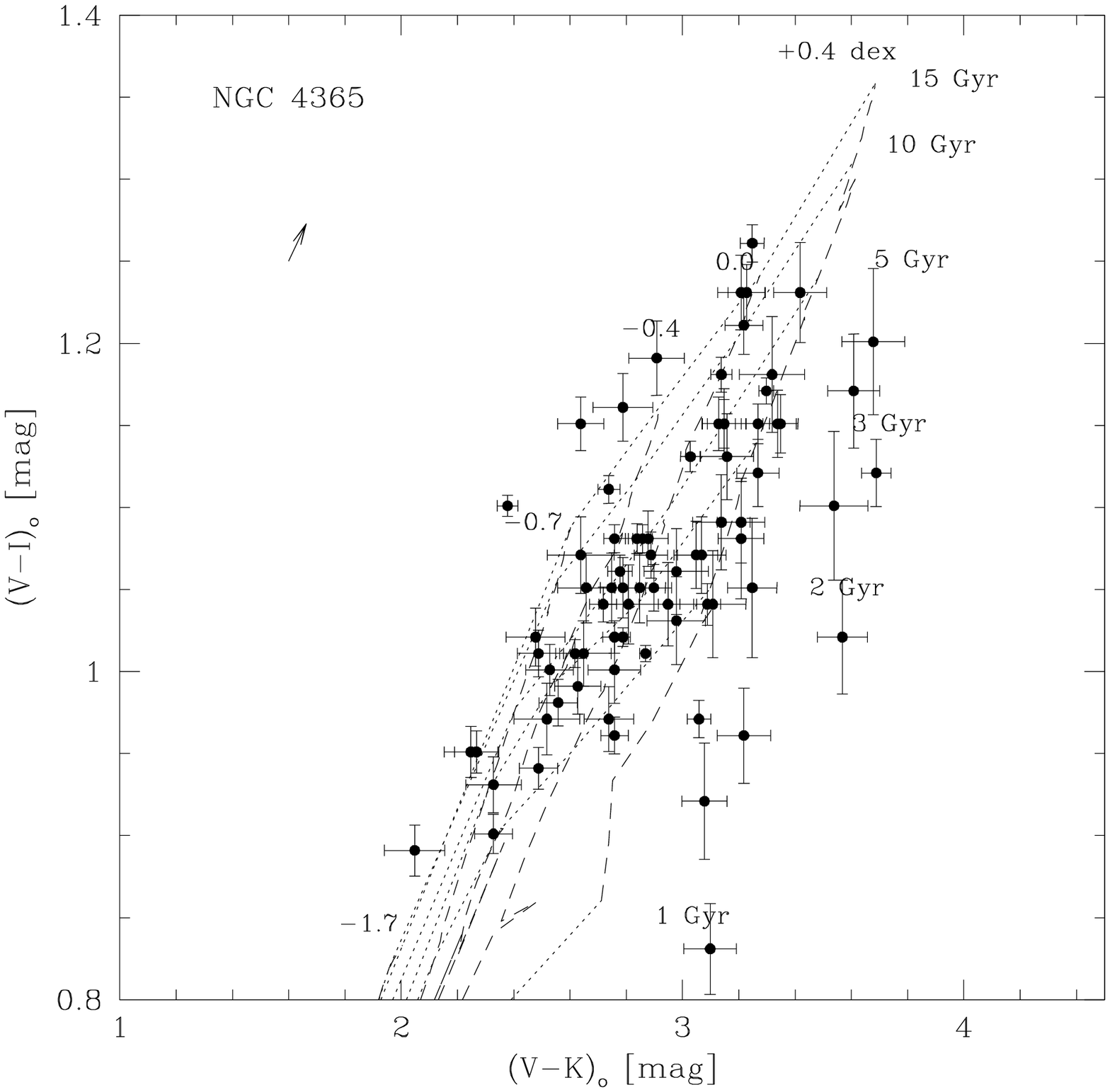}
\end{center}
\caption[]{($V-I$) vs. ($V-K$) diagnostic plots for globular clusters
in NGC 3115 ({\it left panel\,}) and NGC 4365 ({\it right panel\,})
taken from \cite{puzia02}. A significant intermediate-age globular
cluster sub-population is detected in NGC 4365. All used SSP models
indicate that these objects have age between 2 and 8 Gyr and
metallicities $\sim 0.5 Z_\odot-3 Z_\odot$. See contribution of Hempel
et al. (this volume) and \cite{goudfrooij01} for further systems}
\label{ps2}
\end{figure}

\subsection{Colour-Colour Method}
Colour-colour diagrams offer an alternative approach to photometric
ages without the assumption of same underlying globular cluster mass
functions. Combined with a wide wavelength coverage from the optical
to the near-infrared, our group uses the two-colour diagram technique
to derive relative ages of globular clusters in early-type galaxies
(see contribution of Maren Hempel). The advantage of linking optical
and near-infrared colours is that it reduces the age-metallicity
degeneracy very efficiently. The reason for this is the enhanced
metallicity sensitivity of optical/near-infrared colours (e.g. $V-K$)
due to the sampling of metallicity-sensitive giant branch
luminosity. At similar age-sensitivity as purely optical colours
(e.g. $V-I$) this combination minimizes the age-metallicity degeneracy
and increases the total age-sensitivity in a colour-colour diagram.

Figure \ref{ps2} shows the results of an optical/near-infrared study
of two nearby early-type galaxies, NGC~3115 and NGC~4365. Our group
detected a significant population of intermediate-age globular
clusters in NGC~4365 \cite{puzia02} which could not be detected from
optical colours alone. The results were recently confirmed
\cite{larsen02} with spectroscopic methods discussed
below. Counterparts were also found in NGC~1316 \cite{goudfrooij01}
and NGC~5846 (see M.~Hempel et al. in this volume). The subtle but
important point of this study is the fact that an intermediate-age
stellar population can be well-hidden in the diffuse light of
galaxies. No hint for an intermediate-age stellar population in the
diffuse light of NGC~4365 was found so far \cite{davies01}. We clearly
require more near-infrared data on globular cluster systems to
quantify the mass fractions of such intermediate-age globular cluster
sub-populations. Future work \cite{hempel02} could provide such
numbers and strictly constrain the importance of certain galaxy
formation scenarios.

%%%%%%%%%%%%%%%%%%%%%%%%%%%%%%%%%%%%%%%%%%%%%%%%%%%%%%%%%%%%%%%%%%%%%%%%%%%
\section{Spectroscopic Age Dating}

As in the photometric approach, spectroscopic ages are derived from
the comparison of observed parameters with theoretical model
predictions. A commonly-used tool is the so-called diagnostic plot in
which a spectroscopic metallicity tracer is plotted against a
spectroscopic age indicator. The best spectroscopic age indicators for
old stellar populations are Balmer lines. However, there is still no
ideal set or combination of spectroscopic parameters which are
entirely free from the age-metallicity degeneracy. Yet, combined with
a metallicity indicator which is almost independent of age robust
estimates can be achieved. One way to measure the strength of Balmer
lines in distant stellar populations is the use of spectroscopic
indices which do not require high-resolution and high-S/N spectra. The
Lick system \cite{trager98} defines a number of such indices which are
designed to measure the strength of three Balmer lines and many metal
lines.

\subsection{The relatively best Metallicity Indicator}
\begin{figure}[!t]
\begin{center}
\includegraphics[width=4cm]{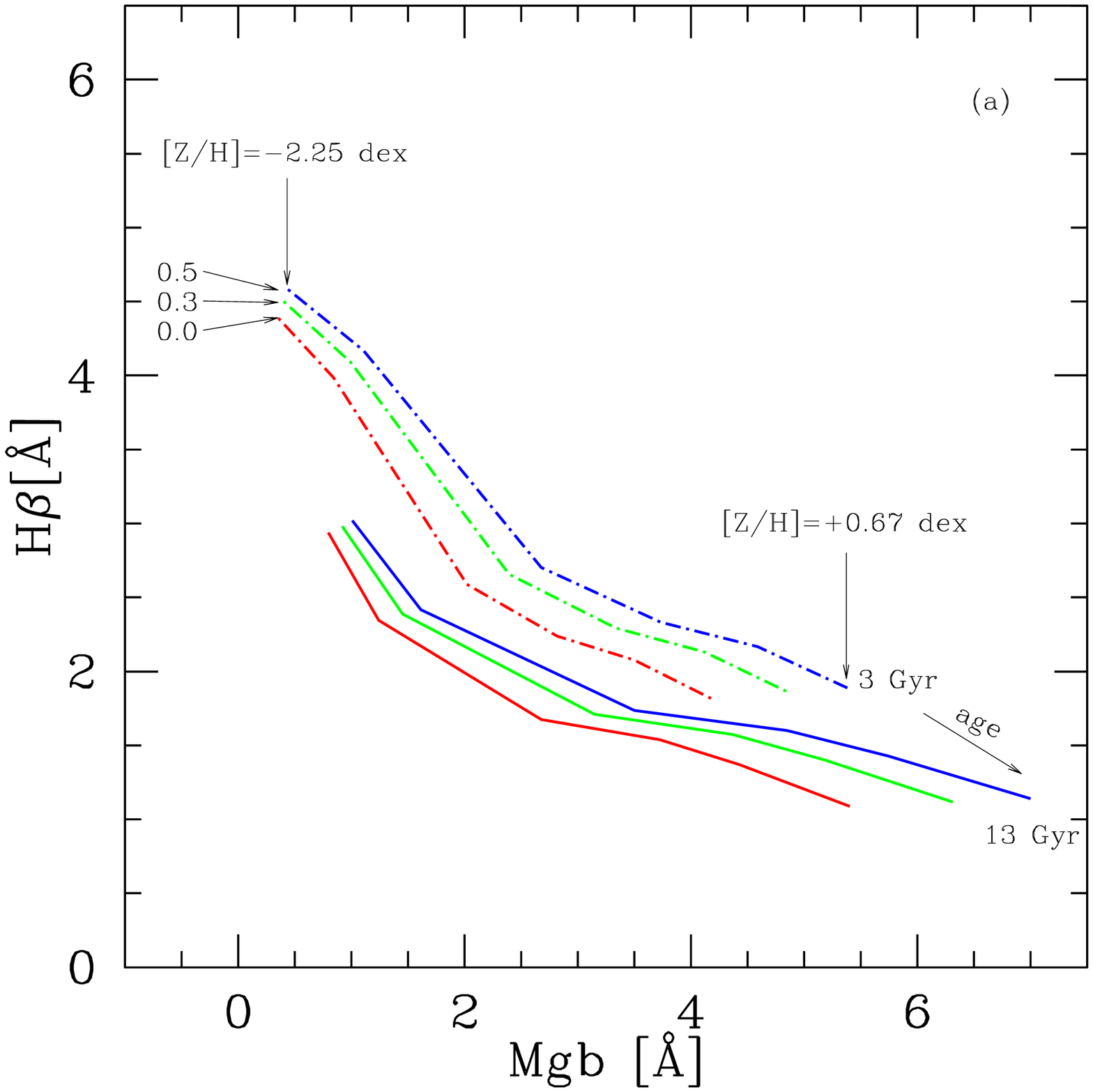}
\includegraphics[width=4cm]{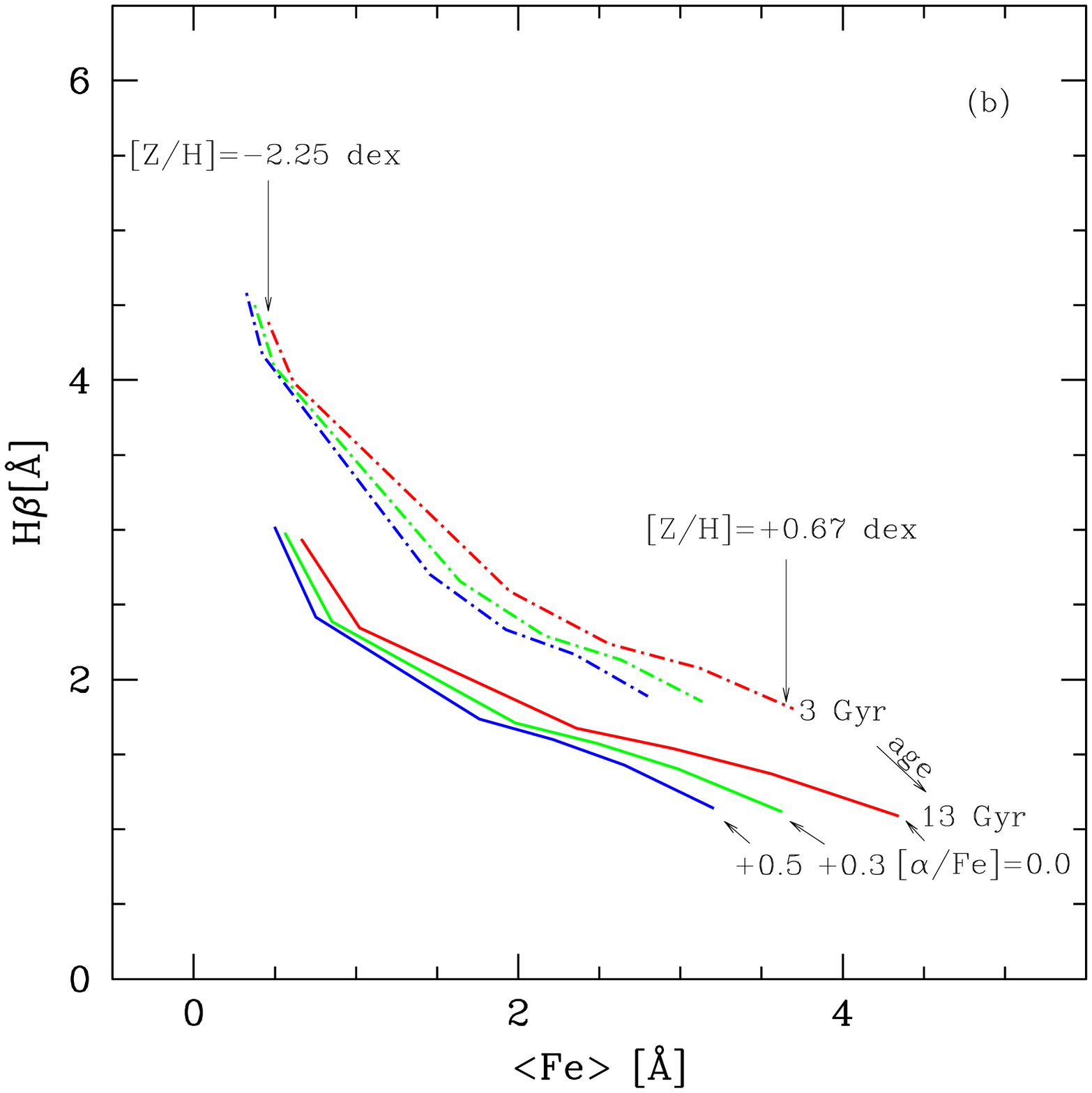}
\includegraphics[width=4cm]{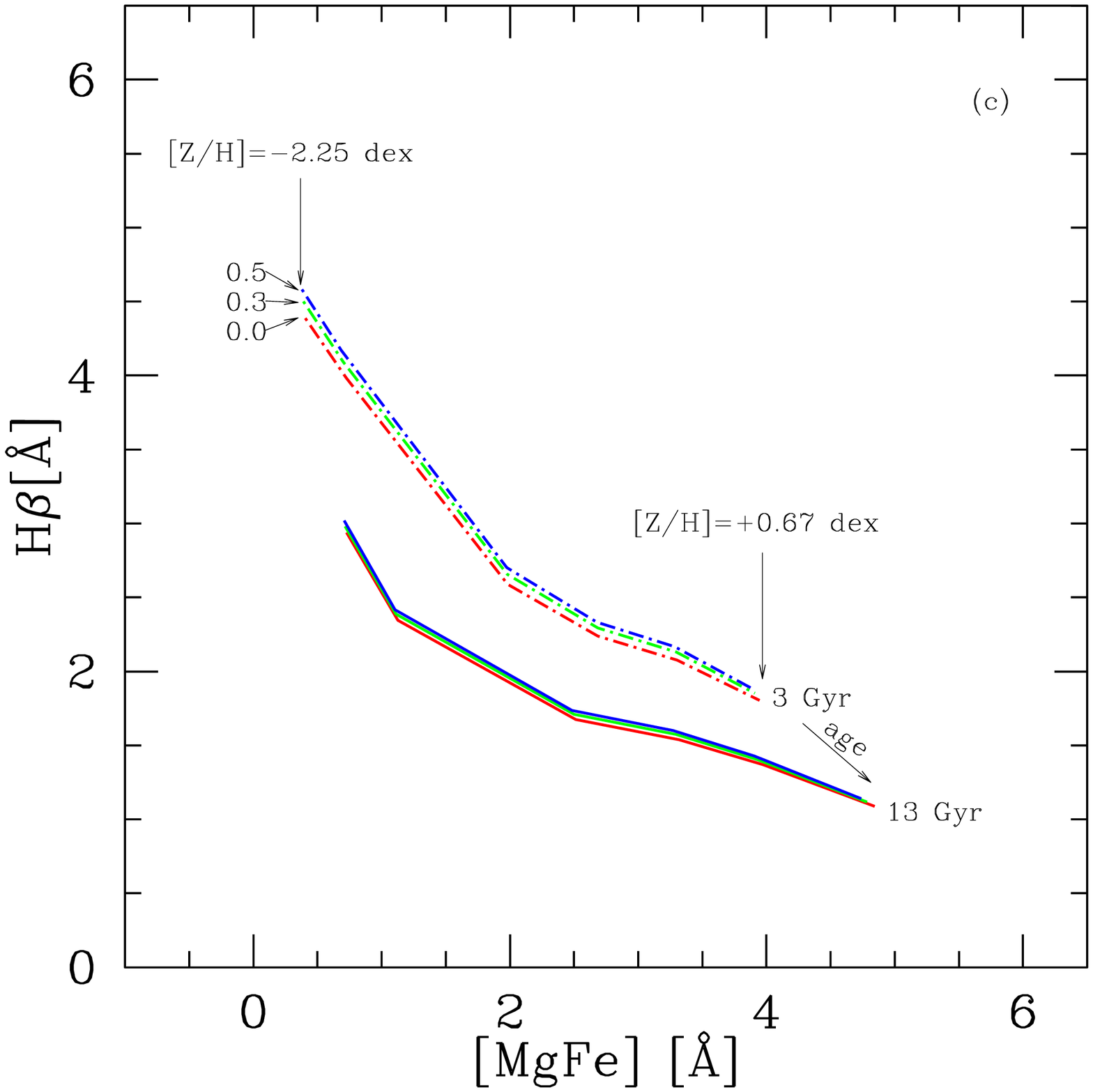}
\end{center}
\caption[]{Two iso-age lines (3 and 13 Gyr) are plotted as a function
metallicity ($-2.25\leq$ [Z/H] $\leq+0.67$) parameterized for three
[$\alpha$/Fe] ratios (0.0, $+0.3$, and $+0.5$ dex) in three different
diagnostic plots. (\textbf{a}): H$\beta$ vs. Mgb. The sensitivity to
[$\alpha$/Fe] can change the age prediction by up to $\sim 5$
Gyr. (\textbf{b}): H$\beta$ vs. $\langle$Fe$\rangle$. Sensitivity to
[$\alpha$/Fe] as in panel a. (\textbf{c}): H$\beta$ vs. [MgFe]. The
isochrones in this diagnostic plot are least affected by [$\alpha$/Fe]
variations. Other Balmer line indices show similar behavior. The
models are taken from \cite{thomas02}}
\label{ps3}
\end{figure}

Depending on the star formation history the chemical composition of a
stellar population is subject to abundance variations, in particular
the [$\alpha$/Fe] ratio \cite{trager00}. These variations blurr the
accuracy of any metallicity diagnostic. New SSP models for stellar
populations with constant [$\alpha$/Fe] ratio \cite{thomas02} allow to
account for varying abundance ratios and to quantify the effect of
changing [$\alpha$/Fe] on spectroscopic indices. A good metallicity
indicator can now be constructed with the requirement of being
insensitive to abundance variations, but simultaneously trace the
total metallicity. Gonzalez \cite{gonzalez93} found discrepant age
predictions of theoretical models depending on the choice of Lick
indices used as metallicity indicators. He speculated that this effect
might be due to variation in the [$\alpha$/Fe] ratio and could be
cured by using the composite index [MgFe]$=\sqrt{{\rm Mg}b\cdot({\rm
Fe5270} + {\rm Fe5335})/2}$. In Figure \ref{ps3} the effects of
changing [$\alpha$/Fe] ratios on old and young isochrones are
shown. Indeed, the composite index [MgFe] is only little affected by
abundance variations and should be used a metallicity indicator in all
spectroscopic diagnostic plots.

\subsection{The relatively best Age Indicator}
The Lick system defines five Balmer line indices, H$\beta$,
H$\gamma_A$, H$\gamma_F$, H$\delta_A$, and H$\delta_F$. The age
sensitivity is different for each index and has never been
quantified. Various combinations of Balmer indices and metallicity
indicators are used in the literature and assigned equal
importance. Here we define a quantity which can be used to determine
the relatively best Lick Balmer-line index to age date stellar
populations given the quality of data and any SSP model. The relative
age-sensitivity of Balmer indices is a function of:
\begin{itemize}
\item $\eta$: mean error of the data
\item $\zeta$: transformation accuracy to the Lick system
\item $\gamma$: mean error of the original Lick spectra
\item $\delta$: accuracy of the Lick fitting functions
\cite{worthey94}, \cite{worthey97}
\item ${\cal D}_{\rm Z}$: dynamic index range at a given metallicity
\item ${\cal S}_{\rm Z, t}$: sensitivity to age and metallicity at a
given metallicity and age (i.e. the impact of the age-metallicity
generacy)
\end{itemize}
where ${\cal S}_{\rm Z,t}$, the degeneracy parameter, is
\begin{equation}
\label{eq:S}
{\cal S}_{\rm Z,t}(I) = \left. \frac{\partial I}{\partial t}
\right|_{\rm Z} \cdot \left. \left( \frac{\partial I}{\partial{\rm
Z}}\right)^{-1}\right|_{t}.
\end{equation}
We define the quantity
\begin{equation}
\label{eq:R}
{\cal R} = \frac{{\cal D}_{\rm Z}\cdot {\cal S}_{\rm
Z,t}}{\sqrt{\eta^2 + \zeta^2 + \gamma^2 + \delta^2}}
\end{equation}
which is maximal for the best age-sensitive index and is essentially
the dynamic scale of an index $I$ at a given age and metallicity
expressed in units of the total uncertainty. Mean values of the
dynamic range $\langle{\cal D}\rangle$ and the degeneracy parameter
$\langle\cal S\rangle$ are used to evaluate ${\cal R}$. Both former
parameters are derived from the SSP models of \cite{maraston02} and
summarised in Table \ref{tab:abcdBalmer}.

We find that from our data the relatively best age diagnostic is the
H$\gamma_{\rm A}$ index followed by the indices H$\delta_A$, H$\beta$,
H$\gamma_F$, and H$\delta_F$. If our data set were infinitely accurate
(i.e. $\eta=0$ in Eq.~\ref{eq:R}) the order of the Balmer-indices
would remain the same. However, ${\cal R}$ is subject to change when
using different SSP models.

\begin{table*}[!t]
\begin{center}
 \caption{Summary of the coefficients in equation \ref{eq:R}. Column 1
        gives the mean data quality of the sample in Figure
        \ref{ps4}. The coefficients in columns 2-8 are given in units
        of \AA. Columns 9-12 are given in dex/Gyr while the unit of
        $\cal R$ in the last column is \AA$\cdot$dex/Gyr. The average
        dynamic age range ${\cal D}_{\rm Z}$ is evaluated at
        [Fe/H]$=-1.35$ and 0.0 between the 1 and 15 Gyr
        isochrone. $\cal S$ is the mean of the degeneracies at two
        different metallicities [Fe/H]$=-1.35$ and 0.0 at two
        different ages 3 and 13 Gyrs.}
\label{tab:abcdBalmer}
\begin{tabular}[angle=0,width=\textwidth]{lp{1.0cm}p{1.0cm}p{1.0cm}p{1.0cm}p{1.0cm}p{1.0cm}p{1.0cm}} 
\hline
\noalign{\smallskip}
 index & $\eta$ & $\zeta$ & $\gamma$ & $\delta$ & 
$\langle{\cal D}\rangle$ & $\langle\cal S\rangle$ & $\cal R$ \\  
\noalign{\smallskip}
\hline
\noalign{\smallskip}
H$\beta$   &  0.60 & 0.232 & 0.22 & 1.30 & 2.62 & 0.274 & 0.488 \\
H$\delta_A$&  0.79 & 1.043 & 0.64 & 1.27 & 7.42 & 0.147 & 0.565 \\
H$\gamma_A$&  0.82 & 0.722 & 0.48 & 1.78 & 9.38 & 0.166 & 0.723 \\
H$\delta_F$&  0.83 & 0.790 & 0.40 & 1.18 & 4.15 & 0.130 & 0.318 \\
H$\gamma_F$&  0.84 & 0.448 & 0.33 & 1.34 & 5.19 & 0.136 & 0.421 \\
\noalign{\smallskip}
\hline
\end{tabular}
\end{center}
\end{table*}

\subsection{Spectroscopic Ages of Globular Clusters in Early-type Galaxies}
\begin{figure}[!t]
\begin{center}
\includegraphics[width=11cm]{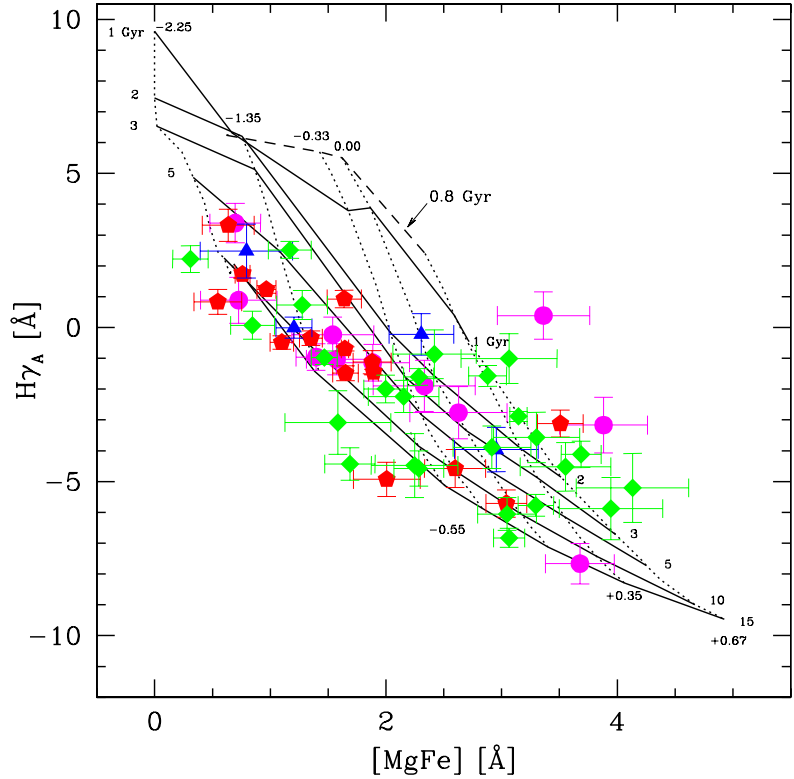}
\end{center}
\caption[]{Globular clusters in the early-type galaxies NGC~1380,
NGC~2434, NGC~3115, and NGC~3379. SSP models were taken from
\cite{maraston02}. Isochrones ({\it solid lines\,}) are plotted for
the ages 1 to 15 Gyr. A very young iso-age line for a stellar
population of 0.8 Gyr is also plotted ({\it dashed
lines\,}). Iso-metallicity lines ({\it dotted lines\,}) are plotted
for values $-2.25\leq$ [Z/H] $\leq+0.67$}
\label{ps4}
\end{figure}

First results of the age dating of globular clusters from our
spectroscopic survey of globular cluster systems in early-type
galaxies are shown in Figure \ref{ps4}. Independent of galaxy,
metal-poor ([Fe/H]$<-0.8$) globular clusters appear to be old and
coeval. Metal-rich ([Fe/H]$>-0.8$) globular clusters, on the other
hand, cover a larger age range and appear on average younger. This
interesting result indicates that metal-poor globular clusters might
have preferentially formed very early, perhaps during or even before
the assembly of their host galaxies. The metal-rich sub-population
appears to form continuously until the recent past. This in turn
implies that only the metal-rich population of clusters is augmented
in later star formation events (see also M.~Kissler-Patig et al. in
this volume and \cite{puzia02b}).

Some caveats have to be addressed when ages are derived from
spectroscopic diagnostic plots. In the metal-poor regime ([Fe/H] $<
-0.8$) at old ages ($t > 10$ Gyr), SSP model tracks start to overlap
due to the increasing importance of hot blue horizontal branch
stars. The consequence are ambiguous age predictions for old
metal-poor populations and a potential age spread at old ages which
cannot be ruled out. In the metal-rich regime ([Fe/H] $>-0.8$) the
accurate age determination might also be hampered by an unknown
contribution of the blue horizontal branch as a consequence of
inaccurate modelling of the mass loss on the giant branch.

\subsection*{acknowledgements}
It is a pleasure to thank my collaborators on various parts of this
work Markus Kissler-Patig, Ralf Bender, Jean Brodie, Paul Goudfrooij,
Maren Hempel, Michael Hilker, Claudia Maraston, Dante Minniti, Tom
Richtler, Roberto Saglia, Daniel Thomas, and Steve Zepf. The financial
support of the German \emph{Deut\-sche For\-schungs\-ge\-mein\-schaft,
DFG\/}, under the grant number Be~1091/10--2 is gratefully
acknowledged.

%INDEX%%%%%%%%%%%%%%%%%%%%%%%%%%%%%%%%%%%%%%%%%%%%%%%%%%%%%%%%%%%%%%%
% Please check with the editor of your book whether he plans to
% include a "mutual" subject index - if so, please code your entries
% in the standard syntax. For your own purposes you may print your
% "personal" index by using the following commands:
%
%\clearpage
%\addcontentsline{toc}{section}{Index}
%\flushbottom
%\printindex
%%%%%%%%%%%%%%%%%%%%%%%%%%%%%%%%%%%%%%%%%%%%%%%%%%%%%%%%%%%%%%%%%%%%%

\end{document}